\documentclass[preprintnumbers,superscriptaddress,pra]{revtex4}
\usepackage{amssymb}
\usepackage{amsmath}
\usepackage{epsfig}
\usepackage{graphicx}
\usepackage{color}

\input{tcilatex}
\begin{document}

\title{Curvature-induced noncommutativity of two different components of
momentum for a particle on a hypersurface}
\author{Q. H. Liu}
\email{quanhuiliu@gmail.com}
\affiliation{School for Theoretical Physics, School of Physics and Electronics, Hunan
University, Changsha 410082, China}
\affiliation{Synergetic Innovation Center for Quantum Effects and Applications (SICQEA),
Hunan Normal University,Changsha 410081, China}
\author{X. Yang}
\affiliation{School for Theoretical Physics, School of Physics and Electronics, Hunan
University, Changsha 410082, China}
\author{Z. Li}
\affiliation{School for Theoretical Physics, School of Physics and Electronics, Hunan
University, Changsha 410082, China}
\date{\today }

\begin{abstract}
As a nonrelativistic particle constrained to remain on an $N-1$ ($N\geq 2$)
dimensional hypersurface embedded in an $N$ dimensional Euclidean space, two
different components $p_{i}$ and $p_{j}$ ($i,j=1,2,3,...N$) of the Cartesian
momentum of the particle are not mutually commutative, and explicitly
commutation relations $[p_{i},p_{j}]\left( \neq 0\right) $ depend on
products of positions and momenta in uncontrollable ways. The \textit{%
generalized} Dupin indicatrix of the hypersurface, a local analysis
technique, is utilized to explore the dependence of the noncommutativity on
the curvatures on a \textit{local point }of the hypersurface. The first
finding is that the noncommutativity can be grouped into two categories; one
is the product of a sectional curvature and the angular momentum, and
another is the product of a principal curvature and the momentum. The second
finding is that, for a small circle lying a \textit{tangential plane}
covering the \textit{local point}, the noncommutativity leads to a rotation
operator and the amount of the rotation is an angle anholonomy; and along
each of the \textit{normal sectional curves} centering the \textit{given
point} the noncommutativity leads to a translation plus an additional
rotation and the amount of the rotation is one half of the tangential angle
change of the arc.
\end{abstract}

\maketitle

\section{Introduction}

In quantum mechanics there are so-called fundamental quantum conditions that
include as the vital part the commutation relations between any pair of
different components of momentum \cite{dirac,dirac1}. The momentum operators
in flat space are well understood, but it is not so in curved space. For a
particle moves on the curved hypersurface that can be modelled as a simple
curved space, the commutation relations for momentum have complicated
structure. Though Dirac proposed in 1950 the standard procedure of
constructing the commutation relations for momentum, and the quantization of
the motion for the particle on the surface has been studied for more than
six decades \cite%
{dirac1,dirac2,dirac3,1968,homma,ikegami,Klauder,kleinert,Golovnev,weinberg,liu11}%
, the understanding of the commutation relations for momentum is still
insufficient, through the curvature-induced effects have been investigated
theoretically and experimentally \cite%
{liu11,liu17,liu13-1,liu13-2,liu13-3,liu14,wang17,liu07,waveguide}. A recent
article contains a mini-review of the current theoretical researches on this
subject \cite{liu19}.

In order to get an unambiguous result on the noncommutativity (\textit{c.f.}
Eq. (\ref{ppd})), let us recall the powerful local analysis in physics and
mathematics. For instance, in the general relativity, the small region of
globally curved spacetime is approximately flat, and a non-linear
differential equation can be made linear one if examining locally. For a
two-dimensional curved surface, the Dupin indicatrix is a standard method
for characterizing the local shape of a surface \cite{lp}, which can be
easily to be generalized to hypersurfaces in higher dimensions to analyze
the local shape \cite{1960}. Such a analysis was performed to investigate
the curvature-induced potential for the particle constrained on the
hypersurface \cite{Golovnev}, yielding a form of the curvature-induced
potential originally predicted by the well-defined \textit{confining
potential formalism} \cite{QP1} (or called thin-layer quantization procedure
\cite{Golovnev}). In present study, the technique is utilized to investigate
the long-lasting noncommutative commutation relations of momentum operators,
revealing novel results which has not been revealed before.

For a nonrelativistic particle constrained to remain on an ($N-1$%
)-dimensional smooth curved surface $\Sigma ^{N-1}$ in flat space $R^{N}$ ($%
N\succeq 2$), one can for the particle define $N$ pairs of Cartesian
variable $(p_{i},x_{i})$ (hereafter $i,j,l=1,2,3,...N$) where $p_{i}$ is $i$%
th cartesian momentum and $x_{i}$ is $i$th coordinate. In classical
mechanics, we know that two different components of the momentum ${\mathbf{p}%
}$ do not commute with each other \cite%
{1968,homma,ikegami,Klauder,kleinert,Golovnev,weinberg},
\begin{equation}
\lbrack p_{i},p_{j}]_{D}=\Pi _{ij}\equiv
\sum_{l=1}^{N}(n_{j}n_{i,l}-n_{i}n_{j,l})p_{l}\neq 0,(i\neq j),  \label{ppd}
\end{equation}%
where subscript $D$ in the square bracket denotes the Dirac bracket, and $%
n_{i}$ is the $i$-th component of the normal vector $\mathbf{n}$ at a point
of the surface $\Sigma ^{N-1}$ and symbol "$,l$" in the subscript stands for
the derivative with respect to the coordinate $x_{l}$, and so forth.

Since 1968 \cite{1968}, there are different approaches to construct the
quantum mechanical commutation relations $[p_{i},p_{j}]=i\hbar \hat{\Pi}_{ij}
$, which has been nevertheless a controversial issue, where $\hat{F}$
denotes the operator form of a classical quantity $F$, and the hat "$\wedge $%
" over the quantity $F$ is usually omitted for convenience. A notoriously
operator-ordering difficulty as to distribute $p_{l}$ in $n_{j}n_{i,l}$ and $%
n_{i}n_{j,l}$ in $\Pi _{ij}$ (\ref{ppd}) is hard to resolve. Take the
distribution problem of inserting $p_{l}$ in $n_{j}n_{i,l}$ for instance,
and there are different approaches. The first approach is a simple
combination of two possibilities $p_{l}n_{j}n_{i,l}$ and $n_{j}n_{i,l}p_{l}$
\cite{homma}, and the second is to consider following four possibilities $%
p_{l}n_{j}n_{i,l}$, $n_{j}p_{l}n_{i,l}$, $n_{i,l}p_{l}n_{j}$, and $%
n_{j}n_{i,l}p_{l}$ \cite{ikegami}. The attempt of Weinberg is to insert $%
p_{l}$ into position-dependent factors forming $n_{j}$ \cite{weinberg},
which is subtle. So far, except for very special case such as the spherical
surface \cite%
{1968,homma,ikegami,Klauder,kleinert,Golovnev,weinberg,liu11,liu17,liu13-1,liu13-2}
and the flat plane, the physical significance of the quantity $\Pi _{ij}$ (%
\ref{ppd}) in general has been an open problem for quite a long time. An
important issue relevant to an nonrelativistic particle constrained on the
hypersurface is that there is the curvature-induced geometric potential \cite%
{liu17}, but we deal with fundamental quantum conditions (\ref{ppd}) which
apply to both nonrelativistic and relativistic case in which there is no
curvature-induced geometric potential \cite{liu19,2016}.

The structure of $\Pi _{ij}$ (\ref{ppd}) for a small area of surface around
a given point is surprisingly simple, which in quantum mechanics becomes
operator-ordering free. It is in sharp contrast to what the noncommutativity
might suggest. Results in section II show that the leading contribution of $%
\Pi _{ij}$ in (\ref{ppd}) can be categorized into two classes. In section
III we construct two geometrically infinitesimal displacement operators
(GIDOs), and demonstrate that these two GIDOs can be divided into two groups
of operators, in which one is purely rotational and another is translational
plus rotational. Section IV presents conclusions and discussions.

\section{A local expansion of the surface equation and noncommutativity
without operator-ordering problem}

Let us consider the surface equation $f(x)=0$, where $f(x)$ is some smooth
function of position $x=(x_{1},x_{2},...x_{N})$ in $R^{N}$, whose normal
vector is $\mathbf{n}\equiv \nabla f(x)/|\nabla f(x)|$. We can always choose
the equation of the surface such that $|\nabla f(x)|=1$, so that $\mathbf{n}%
\equiv \nabla f(x)$. This is because physics does not depend on the specific
form of the surface equations, but depend on the invariants of the surface,
which remain the same for all possible surface equations. Some geometric
invariants include, the normal vector, principal curvatures, and number of
genus, etc.

At any point of the surface, let us put the origin $O$ of an $N$ dimensional
cartesian coordinates at the given point of the surface. In a sufficiently
small region covering the origin $O$, we construct a system of orthogonal
coordinates $(X_{1},X_{2},...X_{N-1},X_{N})$ which can be used to specify a
point in the vicinity of the origin $O(X=0)$ on the hypersurface, and the
surface equation around the origin $O$ can be so chosen $f(X)\equiv
X_{N}-w(X_{1},X_{2},...X_{N-1})$ that $w(X_{1},X_{2},...X_{N-1})$ is Monge's
form of the hypersurface. What is more, we can always choose the coordinates
such that the normal direction $\mathbf{n}$ is along the $X_{N}$-axis and
principal directions are along $N-1$ coordinates $X_{a}$ ($a,b=1,2,3,...N-1$%
), respectively, and the hypersurface is asymptotically represented by the
generalization of the two-dimensional Dupin indicatrix \cite%
{lp,1960,Golovnev},%
\begin{equation}
f(X)=0\longrightarrow X_{N}\approx \frac{1}{2}\sum_{a=1}^{N-1}k_{a}X_{a}^{2},
\label{dupin}
\end{equation}%
where $k_{a}$ is the $a$-th principal curvature of the curve formed by the
intersection of the $X_{a}X_{N}$-plane on the hypersurface $\Sigma ^{N-1}$
at the origin $O$, and the intersections from the normal sections, and there
are in total $N-1$ normal sections. A product $K_{ab}\equiv k_{a}k_{b\text{ }%
}$($a\neq b$) is right the $ab$-th sectional curvature \cite%
{Golovnev,docarmo,mathpage}. The normal vector near the origin $O$ is,
\begin{subequations}
\begin{eqnarray}
\mathbf{n} &\mathbf{\approx }&\nabla
f(x)=(-k_{1}X_{1},-k_{2}X_{2},...,-k_{N-1}X_{N-1},1),\text{or }  \label{nO}
\\
n_{a} &\approx &-k_{a}X_{a},\text{ and \ }n_{N}=1,  \label{n01}
\end{eqnarray}%
which at $O$ reduces to $\mathbf{n}=(0,0,...,0,1)$. The derivative of the
normal vector $\mathbf{n}$ with respect to the coordinate $X_{l}$ gives,
\end{subequations}
\begin{equation}
n_{a,l}\approx \left\{
\begin{array}{c}
k_{a}(1+o(X^{2}))\delta _{l,a} \\
o(X^{2})(1-\delta _{l,a})%
\end{array}%
\right. ,\text{ and }n_{a,N}\approx o(X)\left( 1-\delta _{Na}\right) ,
\label{nal}
\end{equation}%
where $o(X)$ and $o(X^{2})$ denote quantities of order $X$ and $X^{2}$,
respectively. At $O$, we have, respectively, the mean curvature,
\begin{equation}
M\equiv {-}\sum_{i=1}^{N}n_{i,i}=\sum_{a=1}^{N-1}k_{a},
\label{meancurvature}
\end{equation}%
and,
\begin{equation}
\sum_{i,j=1}^{N}\left( n_{i,j}\right) ^{2}=\sum_{a=1}^{N-1}k_{a}^{2}.
\label{k2}
\end{equation}%
The central results of the present study are, up to the leading term,%
\begin{equation}
\lbrack
p_{a},p_{b}]_{D}=\sum_{l=1}^{N}(n_{b}n_{a,l}-n_{a}n_{b,l})p_{l}\approx
-K_{ab}L_{ab},(a\neq b),  \label{local}
\end{equation}%
where $L_{ab}\equiv X_{a}p_{b}-X_{b}p_{a}$, and,
\begin{equation}
\lbrack p_{a},p_{N}]_{D}\approx -k_{a}p_{a}.  \label{angle}
\end{equation}%
In consequence, we have the \textit{local} commutation relations in quantum
mechanics,
\begin{equation}
\lbrack p_{a},p_{b}]\approx -i\hbar K_{ab}L_{ab},\text{ and }%
[p_{a},p_{N}]\approx -i\hbar k_{a}p_{a}.  \label{qlocal}
\end{equation}%
These two sets of commutation relations are remarkable for they are free
from operator-ordering difficulty.

Two immediate remarks on these local relations (\ref{angle}) and (\ref{local}%
) follow. 1) They depend on the local\ geometric invariants of the surface
such as $K_{ab}$, $k_{a}$, $L_{ab}$ and $p_{a}$ etc.,\ so they hold
irrespective of coordinates chosen. 2) The brackets (\ref{local}) and (\ref%
{angle}) are zero once $K_{ab}$ and $k_{a}$ are zero respectively, as
expected.

\section{Geometrically infinitesimal displacement operators and rotations}

Now we further investigate the physical significances of the commutation
relations (\ref{qlocal}).

First, we construct a GIDO along a small circle which is approximated by%
\textit{\ a small square} in the \emph{tangential} $X_{a}X_{b}$-plane around
the origin $O$; and let the small square be formed by four points at A($%
-\delta X_{a}/2,-\delta X_{b}/2$), B($\delta X_{a}/2,-\delta X_{b}/2$), C($%
\delta X_{a}/2,\delta X_{b}/2$) and D($\delta X_{a}/2,-\delta X_{b}/2$),
with center at the origin $O$ with $\left\vert \delta X_{a}\right\vert
=\left\vert \delta X_{b}\right\vert $. The initial and final points of the
displacements coincide at point A($-\delta X_{a}/2,-\delta X_{b}/2$), and
order of the displacement is A$\rightarrow $B$\rightarrow $C$\rightarrow $D$%
\rightarrow $A. We have a GIDO along a small square $\square $ABCD,
\begin{equation}
G_{\square }\equiv e^{i\frac{\delta X_{b}p_{b}}{\hbar }}e^{i\frac{\delta
X_{a}p_{a}}{\hbar }}e^{-i\frac{\delta X_{b}p_{b}}{\hbar }}e^{-i\frac{\delta
X_{a}p_{a}}{\hbar }}\approx e^{\frac{\delta X_{a}\delta X_{b}}{\hbar ^{2}}%
[p_{a},p_{b}]}\approx e^{-\frac{i}{\hbar }\left( \delta X_{a}\delta
X_{b}K_{ab}\right) L_{ab}}.  \label{ro}
\end{equation}%
In calculation, the Baker-Campbell-Hausdorff formula for two possibly
noncommutative operators $u$ and $v$ as $e^{u}e^{v}\approx
e^{u+v}e^{[u,v]/2} $ is used. We see that the GIDO $G_{\square }$ (\ref{ro})
is a rotational operator on the $X_{a}X_{b}$-plane, and the angle of the
rotation is $\left( \delta X_{a}\delta X_{b}K_{ab}\right) $ which is the
sectional anholonomy. It is originally defined by the angle of rotation of
the vector as it is accumulated during parallel transport of the vector on a
the hypersurface along the the small circle on the $X_{a}X_{b}$-plane. The
angle anholonomy formed by a loop covering an finite area $\Delta S$ on the
hypersurface is given by,
\begin{equation}
\sum_{a,b=1}^{N-1}\int_{\Delta S}K_{ab}dX_{a}\wedge dX_{b},  \label{anh}
\end{equation}%
where the finite area $\Delta S$ is formed by infinitely many flat pieces
covering the area, and $\sum_{a,b=1}^{N-1}\oint K_{ab}dX_{a}\wedge
dX_{b}=2\pi \chi $, where $\chi $ is the Chern number.

If the hypersurface is a two-dimensional spherical surface, the angle
anholonomy is equal to the solid angle subtended by loop. If the surface is
locally a saddle, the infinitesimal angle anholonomy is negative. If it is a
cylinder whose gaussian curvature is vanishing, the angle anholonomy is zero.

Secondly, considering \textit{the small arc length} from E$(-\delta
X_{a},-\delta X_{N})$ via $\mathit{O}$ to G$(\delta X_{a},-\delta X_{N})$
along the small portion of the normal sectional curve on the \emph{normal} $%
X_{a}X_{N}$-plane at the origin, we immediately find that the commutator $%
[p_{a},p_{N}]\approx -i\hbar k_{a}p_{a}$ leads to a displacement plus an
additional rotation. To see it, we construct following GIDO which shifts a
quantum state along the arc from point E $\rightarrow $ $O$ $\rightarrow $
G,
\begin{equation}
G_{\frown }\equiv \exp \left( -i\frac{\delta X_{a}p_{a}-\delta X_{N}p_{N}}{%
\hbar }\right) \exp \left( -i\frac{\delta X_{a}p_{a}+\delta X_{N}p_{N}}{%
\hbar }\right) \approx \exp \left( -i\frac{2\delta X_{a}p_{a}}{\hbar }%
\right) \exp \left( -\frac{\delta X_{a}\delta X_{N}}{\hbar ^{2}}\left[
p_{a},p_{N}\right] \right) .  \label{last-2}
\end{equation}%
In right-handed side of this equation, we see two parts, and one is a simple
translational operator $\exp \left( -i\frac{2\delta X_{a}p_{a}}{\hbar }%
\right) $ and another is,
\begin{equation}
\exp \left( -\frac{\delta X_{a}\delta X_{N}}{\hbar ^{2}}\left[ p_{a},p_{N}%
\right] \right) =\exp \left( i\frac{\delta X_{a}\delta X_{N}}{\hbar }%
k_{a}p_{a}\right) .  \label{last-1}
\end{equation}%
The physical significance becomes evident. The arc length element of along E
$\rightarrow $ O $\rightarrow $ G is $ds\equiv 2\sqrt{\delta
X_{N}^{2}+\delta X_{a}^{2}}\approx 2\delta X_{a}$ with noting that $\delta
X_{N}\approx k_{a}X_{a}\delta X_{a}=o(X)\delta X_{a}$ from (\ref{dupin}).
The change of the tangential vector along the arc is $-\delta \theta \equiv
k_{a}ds\equiv 2k_{a}\sqrt{\delta X_{N}^{2}+\delta X_{a}^{2}}\approx
2k_{a}\delta X_{a}$, and we have from above equation (\ref{last-1}),%
\begin{equation}
\exp \left( i\frac{\delta X_{a}\delta X_{N}}{\hbar }k_{a}p_{a}\right)
\approx \exp \left( -\frac{i}{\hbar }\left( \frac{-\delta \theta }{2}\right)
L_{Na}\right) ,  \label{last}
\end{equation}%
where an angular momentum operator defined by a torque of momentum $p_{a}$
with respective to point ($0,-\delta X_{N}$) is $L_{Na}\equiv \delta
X_{N}p_{a}$.

Let us move a quantum state along closed curves formed by piecewise smooth
normal sectional lines, the rotation operator gives an accumulation of the
rotational angle is $\sum \delta \theta =2\pi $. Specially, when the surface
is a two-dimensional spherical surface, the normal sectional curves are
great circles and the GIDO $G_{\frown }$ for a great circle leads to that
the total angular change is $2\pi $.

Thus, we have demonstrated that two seemingly different kinds of
noncommutativity, given by (\ref{qlocal}), have the same crucial parts:
rotation operators given by\ $G_{\square }$ (\ref{ro}) and (\ref{last}) in $%
G_{\frown }$ (\ref{last-2}), respectively. The amount of the rotations
depends on the curvature of the surface.

\section{Conclusions and discussions}

For a nonrelativistic particle constrained to remain on a hypersurface,
Dirac brackets for two different components of momentum are not mutually
commuting with each other. The noncommutativity $\Pi _{ij}$\ on a local
point of the hypersurface is examined and results show that the
noncommutativity is due to the local curvature of the surface. At the point,
there are, respectively, $(N-1)(N-2)/2$ mutually perpendicular
two-dimensional \emph{tangential planes} and $N-1$ mutually perpendicular
\emph{normal sectional curves}. In quantum mechanics, with GIDOs constructed
on the base of the noncommutativity, we find that, at the point, for a small
circle lying on each of the tangential planes covering the point the
noncommutativity leads to a rotation operator and the amount of the rotation
is an angle anholonomy, and for a short arc length along each of the
intersecting curves\ centering the given point the noncommutativity leads to
a translation plus an additional rotation and the amount of the rotation is
one half of the tangential angle change of the arc. All results are obtained
by examination of the noncommutativity, without necessarily knowing the form
of the momentum.

In many aspects our results are in sharp contrast to what the intuition
suggests. For instance, the locally approximated flatness of the surface
suggests that the momentum might reduce to the usual one, but it is not the
case for that the noncommutativity depends on the curvature. The
noncommutativity of commutation relations for momentum operators on a local
point remains, but the heavy operator-ordering difficulty is got rid of.
There is no angular momentum operator in the commutation relations $%
[p_{a},p_{N}]\approx -i\hbar k_{a}p_{a}$, but they can certainly have
quantum states on the surface angularly shifted.

\begin{acknowledgments}
This work is financially supported by National Natural Science Foundation of
China under Grant No. 11675051.
\end{acknowledgments}

\end{document}